\input preprint.sty
\input epsf
\pptstyle
\title{Red--shifts near black holes}

\author{Adam D Helfer}

\address{Department of Mathematics, Mathematical Sciences Building, 
University of Missouri, Columbia, Missouri 65211, U.S.A.}

\jl{6}

\beginabstract
A simple ordinary differential equation is derived governing the red--shifts of
wave--fronts propagating through a non--stationary spherically symmetric
space--time.  Approach to an event horizon corresponds to approach to a fixed
point; in general, the phase portrait of the equation illuminates the
qualitative features of the geometry.  In particular, the asymptotics of the
red--shift as a horizon is approached, a critical ingredient of Hawking's
prediction of radiation from black holes, are easily brought out.  This
asympotic behavior has elements in common with the universal behavior near
phase transitions in statistical physics.  The validity of the Unruh vacuum for
the Hawking process can be understood in terms of this universality.
The concept of surface gravity is
extended to to non--stationary spherically symmetric black holes.
Finally, it is shown that in the non--stationary case, Hawking's
predicted flux of radiation from a black hole would be modified.
\endabstract


\date



\def\const{{\rm const}\, {}}
\def\lesssim{{\mathrel{<\atop\sim}}}
\def\gtrsim{{\mathrel{>\atop\sim}}}
\def\CMP{{Commun. Math. Phys.}}
\def\scri{{\cal I}}
\newcount\EEK
\EEK=0
\def\eek{\global\advance\EEK by 1\eqno(\the\EEK )}
\newcount\FEET
\FEET=0
\def\fnote#1{\global\advance\FEET by 1\footnote{${}^{\the\FEET}$}{#1}}

\section{Introduction}

By definition, a black hole cannot be directly observed by someone in
its exterior.  For this reason, it is the geometry of the \it neighborhood \rm
of the horizon, as measured by later and later observers, which is of interest
in the external physics of the holes.  A basic element of this geometry is
ray--tracing:  looking in from far away towards where the hole will form (and
neglecting opacity), what would one see?

An estimate of just this, for rays sufficiently close to the horizon
in the Schwarzschild (or more generally Kerr--Newman\fnote{In what
follows, it will be understood that the Reissner--Nordstrom and
Kerr solutions are special cases of Kerr--Newman.}) geometries, is a
cornerstone of Hawking's (1974, 1975) prediction of thermal radiation from black
holes.  In the Schwarzschild case, Hawking argued that the asymptotic 
ray--tracing relation between surfaces of constant
phase labeled by retarded time $u$ in the future to those of constant phase
labeled by advanced time $v$ in the past would be
$$v(u)\simeq C\exp -u/(4m)\, ,\eek$$\xdef\adelbert{\the\EEK}%
as $u\to +\infty$, that is, as the horizon is approached.  Here $m$ is the mass
in geometric units.  It is $v(u)$ which is the key function in the Hawking
analysis; for example,the Hawking temperature 
comes out of the calculation as $T_{\rm H}= -(1/2\pi )\ddot v /\dot v$.

The present paper is concerned with the justification of the
relation~(\adelbert ), and the derivation of similar relations for more general
situations.  It will be well to first explain what is known.

Hawking's argument for the relation~(\adelbert ) was brief and perhaps
elliptic; the taking into account of the effects of tracing
the rays back through the collapsing matter was not really spelled out.  
The argument relied on specific properties of the Schwarzschild
(or Kerr--Newman) solutions.  A fuller treatment of the spherically symmetric
case, assuming a static exterior, was given by Birrell and Davies (1982).  Their
argument was by computation in a particular coordinate system, and its relation
to invariant geometry is not manifest.

These arguments leave open two sorts of issues.  First, one would like an
invariant explanation of where the relation (\adelbert ) comes from; and
second, one would like to be able to treat a broader class of objects.  For
example, one would like to know what happens for a black hole accompanied by an
accretion disk, or a spherically symmetric hole by a (non--static) nebula.  
What happens if the collapsing matter is not sharply bounded (so the event
horizon may not be in a static exterior)?\fnote{There is a loose sense in which
many astrophysical black holes are expected to be Kerr--Newman (``black hole
uniqueness'').  However, this is because for this class of holes
one \it assumes \rm that the hole
approaches a stationary state and is isolated (that is, has only the
electromagnetic stress--energy in its exterior).}
If we are concerned
(as we are in the case of the Hawking process) with a star that has contracted
to within atomic or even nuclear dimensions of its Schwarzschild radius, are we
justified in treating it as a classical bounded object? And what about bodies
which are not quite black holes?  What happens if a body collapses to
$r=2m(1+\epsilon )$ and stays there?   To investigate conceptual questions
about the Hawking process, it is desirable to have an invariant way of
understanding and treating these issues.

The aims of the present paper are: (a) to bring out very clearly where the
formula~(\adelbert) comes from; (b) to show that it has a flavor of
``universal'' behavior in the sense of statistical physics; (c) to show that a
similar result will hold without any assumption of asymptotic stationarity or
vacuum exterior for the hole; and (d) to present a general framework for the
study of similar questions.

In the course of this analysis, we clarify some general features of the
structure of black holes.  We shall find a way of extending the concept of
surface gravity to non--stationary spherically symmetric black holes.  This new
measure of surface gravity will be a global one, determined by red--shifts of
light rays passing close to the horizon.  While global, it is given by a
well--defined integral formula in terms of the local geometry.   This enables
us to ask which details of the geometry the surface gravity is sensitive to: 
which contribute to the integral?

We shall see that, as the hole is approached, all contributions from the regime
prior to the formation of the event horizon are suppressed, and so the surface
gravity is independent of the details of the formation of the hole, as one
would expect.

We shall also see that the surface gravity is essentially insensitive to the
geometry of the horizon!  This is in apparent conflict with the usual view that
the surface gravity is determined by the gradient of the norm of a Killing
vector at the horizon.  There is no real conflict, however.  The usual
definition of the surface gravity is really a global one, since Killing's
equation serves to evolve the Killing field throughout the space--time, and the
Killing field must be normalized at infinity.  Thus Killing's equation provides
a sort of rigidity which connects data in one part of the space--time to data
in another.  This rigidity makes it impossible to use the Killing field to
localize contributions to the surface gravity.  Our formula, on the other hand,
expresses it as an integral, and one can say which elements contribute more
substantially than others. What the surface gravity turns out to depend on is
how the neighborhood of the horizon is linked to future null infinity, not on
what happens at the horizon itself.

Of course, in the stationary case, the understanding of the surface gravity
developed here, and the usual one, are simply alternative ways of regarding the
same thing.  In the non--stationary case, we have a new candidate definition
for the surface gravity.  One would not expect this to be constant.  For
example, if a black hole were enclosed in a shell of matter (or radius far
larger than the gravitational radius of the hole), one would expect all physics
within the shell to be red--shifted due to the shell's potential.  Thus the
surface gravity of the hole should appear smaller from outside the shell than
if the shell were not there.  If the shell were moved slowly
(``adiabatically''), one would expect to a time--dependent surface gravity,
with time--dependence given by the time--dependent red--shift due to the
varying potential of the shell.

Our formula for the surface gravity is well--defined in such circumstances, 
and beyond the adiabatic approximation.  It
remains independent of the details of the formation of the hole, and (in the
sense of the second paragraph previous) of the geometry of the horizon.

The techniques we use are fairly elementary ones from differential geometry and
ordinary differential equations.  We shall show that the function $v(u)$ is
governed by a certain system of ordinary differential equations. (These
equations are equivalent to some of the familiar optical equations.)  Approach
to an event horizon corresponds to approach to a certain kind of fixed point
in the phase diagram; the asymptotics of the equations near the fixed point are
what lead to the constancy of the surface gravity.  

Most of this paper is concerned with the spherically symmetric case.   Section
2 describes the universal character of the asymptotic behavior of the
red--shift. Section 3 derives the main equations governing the evolution of the
red--shift, and Section 4 discusses their consequences, in particular, the
behavior of red--shifts near an event horizon.  Section 5 contains comments on
the non--spherically symmetric case.  Section 6 sketches the implications for
Hawking's proposal in the case of non--stationary black holes.

\it Conventions.  \rm  The conventions are those of Penrose and Rindler
(1984--6).  The metric has signature $+{}-{}-{}-$, and the curvature satisfies
$\nabla _a\nabla _b v^d-\nabla _b\nabla _av^d=R_{abc}{}^d v^c$.  We use natural
units.

\section{The Universality}

I will begin with a simple physical picture which brings out the ``universal''
character of the red--shift, without actually establishing its functional form.
(Precisely what is meant by this will appear shortly.) This argument 
contains some restrictive assumptions, but its simplicity makes it worthwhile.

Consider a spherically symmetric collapsing object in general relativity.  We
shall assume the object has a sharp boundary, or limb, and that exterior to
this is vacuum.  Then the exterior will be a portion of the Schwarzschild
solution.  Inside the collapsing object is another metric, the details of which
will be unimportant.  We will only need to assume that the interior metric
joins to the exterior metric suitably smoothly.  (A metric of class $C^1$ would
be more than enough.)  We shall assume a black hole forms, so that as $t\to
+\infty$  the coordinate $r(t)$ of the limb will satisfy $r(t)\to 2m$.

Now trace backwards in time a radially--directed wave--packet.  The wave packet
in the distant future (near future null infinity, $\scri ^+$), has some
characeristic wave--length with respect to the frame defined by the spherical
symmetry at $\scri ^+$.   We are going to follow it backwards in time, through
the collapsing object, and out to the distant past ($\scri ^-$).  We do this in
two stages, which have different physical significances.

\epsfbox{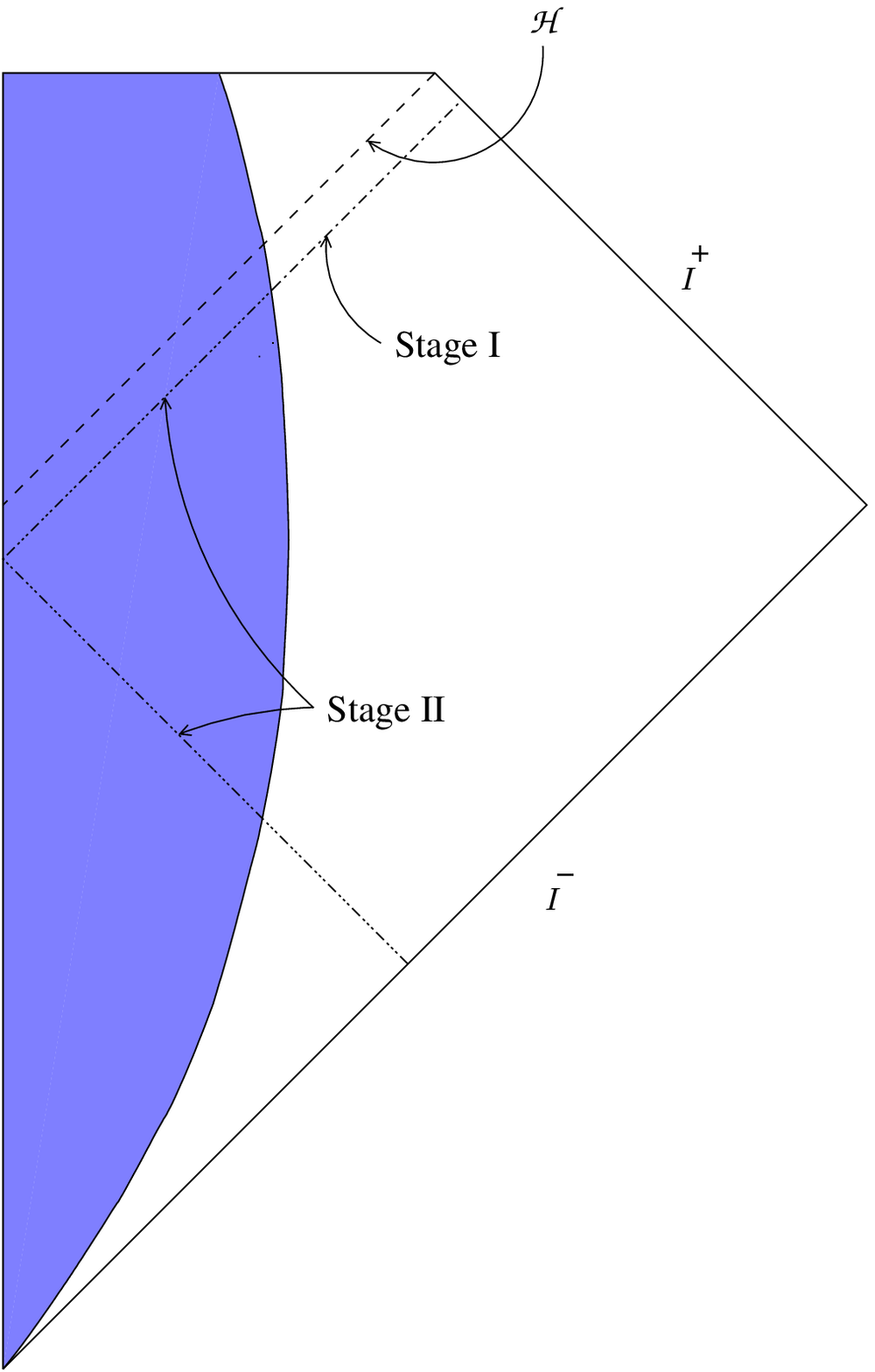}
\figure{Penrose diagram for the spherically symmetric collapse of a massive
body (shaded).  The event horizon is ${\cal H}$.  A geodesic slightly before
${\cal H}$ is also shown, divided into two stages along which we consider the
propagation of a wave packet.}

Stage I.  This covers tracing the packet back from $\scri ^+$ to the limb of
the collapsing object, which is (say) at coordinate $r(t_{\rm packet\ out})$.
If $r(t_{\rm packet\ out})\simeq 2m$, then the wave packet is deep in the
potential well and has been very red--shifted.  The precise measurement of the
red--shift depends on the choice of the observer's frame at $r(t_{\rm packet\
out})$.  However, relative to, say, observers freely falling into the hole, the
red--shift is large, and becomes larger as later and later packets are used, 
since for these $r(t_{\rm packet\ out})\to 2m$.  As later and later
packets are used, this portion of the trajectory corresponds to a divergent
red--shift.

Stage II.  This covers tracing the packet back through the collapsing object,
out the other side, and off to past null infinity $\scri ^-$.  We would expect
this propagation to, to some extent, undo the red--shift of 
the journey from $\scri
^+$ to the limb.  However, it cannot undo all of the red--shift, in the
following sense.  \it  Throughout the second stage of the trip, there is no
propagation which can give rise, even in the limit of later and later times, to
an infinite red--shift.  \rm  On this portion of the trip, the red--shift as
the packet is propagated through the collapsing body remains finite, since this
is only propagation over a finite non--singular region.  Also the propagation
from the collapsing body to $\scri ^-$ introduces no divergent red--shift,
since on this segment of the trip, the radius of the body stays bounded away
from $2m$ so no infinitely deep potential well is involved.

The total red--shift is the product of the red--shifts for Stages I and II, and
thus has the form of a part which diverges (as later and later packets are used)
times a part which remains regular.  The divergent part depends only on the
propagation through the vacuum region exterior to the body, and so is
independent of the details of the collapsing body, that is, universal for the
class of spherical sharply--bounded bodies of given mass.  

We may extract the invariant information in this universality class as
follows.  Write the  red--shift factor as $\dot v  (u)=w_{\rm I}w_{\rm II}$ for
the product of the two stages; then, since $w_{\rm II}$ will tend to a
constant, we see that the limiting behavior of  $(\log\dot v)\dot{} =\ddot v
/\dot v$ is independent of the trip in Stage II.  It is also independent of the
particular choice of Lorentz frame for measuring red--shifts near $r=2m$, since
changes of frame will only contribute smooth multiplicative factors with finite
limits.  At the present level of analysis, the limiting form of  $(\log\dot
v)\dot{}$ could in principle be anything of the form $f(u/m)/m$, where $u$ is
the retarded time and $f$ is a dimensionless function.  Explicit computations
show that $-(\log\dot v)\dot{}\to \kappa =1/4m$, the surface gravity of the
black hole.

While we have emphasized that there is an element of physics here in
common with the universal behavior familiar in statistical mechanics, it would
be wrong to think that the parallel extends, in any obvious way, very much
further.  The picture drawn here of the formation of a black hole differs from
that of a conventional thermodynamic system fundamentally.  First, we are
concerned here with an essentially non--stationary process, since the
time--dependence of the limb's trajectory is crucial.  Second, the space--time
is not a homogeneous extensive system.  For example, one would not have a
correlation length in the usual sense, although at a formal level one can treat
the red--shift factor $\dot v$ as such.

\section{The Case of Spherical Symmetry}

In this section we derive the differential equations governing the evolution of
the red--shift, in the case of spherically symmetric space--times.

We shall assume $(M,g_{ab})$ is a spherically symmetric space--time, which is
asymptotically flat at past and future null infinity in a sense strong enough
to guarantee the existence of $\scri =\scri ^+\cup \scri ^-$ with their usual
differential--geometric structures.  Choose Bondi systems $(u,\theta ,\varphi
)$ on $\scri ^+$ and $(v,\theta,\varphi )$ on $\scri ^-$  respecting the
spherical symmetry.   Then $u$ will be called the retarded time coordinate, and
$v$ the advanced time coordinate.  Fix too a null vector field $n^a$ tangent to
the generators of $\scri ^+$ and a null covector field $l_a$ orthogonal to the
Bondi cuts at $\scri ^+$, with $l_an^a=1$.  The normalization $n^a\nabla _a
u=1$ is assumed.

\it Caution.  \rm  In most papers it is conventional to use $n^a$ for a tangent
vector to $\scri ^+$ and $l^a$ for a tangent vector to $\scri ^-$.  We shall \it
not \rm do this.  The arguments here are global ones involving parallel
transport between $\scri ^-$ and $\scri ^+$, and it is most natural to take
$n^a$ and $l^a$ defined at $\scri ^+$ and then extend them to $\scri ^-$ by
transporting along certain paths.  This gives the opposite--to--usual senses
for the vectors on $\scri ^-$.

\it Note.  \rm  A rigorous treatment of the differential geometry at $\scri$
would involve conformally rescaling the metric and also rescaling the
components of tangent vectors so that their limits took finite values on
$\scri$.  This seemed overly formal for the relatively simple geometric issues
to be considered here.  Thus all calculations in this paper ``at $\scri$''
should be understood as done close to $\scri$, in the asymptotically flat
regime.  The sophisticated reader will have no difficulty recasting the results
and arguments here in a more formal mold.

\subsection{Parallelism at Infinity}

One might hope that very far away from an isolated system one can use parallel
transport to provide an essentially unambiguous way of identifying vectors at
one point with vectors at another.  In general, the situation is not quite so
simple (because, while the gravitational field is locally weak, the regime far
away from the source contains points very distant from each other). It turns
out  however that this \it is \rm a reasonable hypothesis in the spherically
symmetric case.  (It is gravitational radiation which interferes with the use
of aparallel transport to identify vectors at different points of $\scri$; in a
spherically symmetric space--time, there is no gravitational radiation.) In
other words, we assume that over $\scri =\scri ^+\cup \scri ^-$ we have a
natural way of identifying vectors at different points, which we call a
parallelism. 

One important conseqence of this is that the red--shift of a radial null
geodesic is well--defined.  Suppose that we send in a wave packet with
wave--vector $l^a_{\rm in}$ in the distant past (at $\scri ^-$), and it emerges
as $l^a_{\rm out}$ at $\scri ^+$.  In order to compare $l^a_{\rm in}$ and
$l^a_{\rm out}$ we must have some way of identifying the tangent spaces at the
point of emission on $\scri ^-$ and the point of arrival on $\scri ^+$.  It is
this which the parallelism provides.  Using the parallelism, we can regard
$l^a_{\rm in}$ and $l^a_{\rm out}$ as elements of the same space.  Then the
symmetry of the situation implies that that must be multiples of each other,
and the red--shift is simply the constant of proportionality.

It should be noted that the red--shift factor can be given directly. Take a
$u=\const$ surface near $\scri ^+$ and trace it backwards until it emerges as
$v=v(u)=\const$ near $\scri ^-$.  Then the function $v=v(u)$ is the mapping of
surfaces of constant phase, and $\dot v (u)$ is precisely the red--shift
factor.\fnote{While it may seem that $\dot v (u)$ gives a definition of the
red--shift independent of the existence of a parallelism, this is not really
so.  For in order to \it interpret \rm $\dot v (u)$ as a red--shift, one needs
to know how to relate the clocks near $\scri ^+$ (relative to which the
coordinate $u$ is normalized) to those near $\scri ^-$ (relative to which $v$
is normalized.)}

\subsection{Holonomy}

In this subsection, we derive one of the two main equations for the red shift.
This equation arises from a simple holonomy argument, and at first seems to be
all that is necessary for understanding the red--shift.  We shall see, however,
in the next subsection, that there is a subtlety, and this equation must be
supplemented by another in order to understand the black--hole state.

The red--shift arises by comparing the result of propagating light rays through
space--time from $\scri ^-$ to $\scri ^+$ with the parallelism at infinity. 
Mathematically, this corresponds to a holonomy, that is, parallel propagation
around a certain closed path $\Gamma _u$.  In fact, it will be convenient to 
begin by defining the oppositely oriented path $-\Gamma _u$, as follows.

\epsfbox{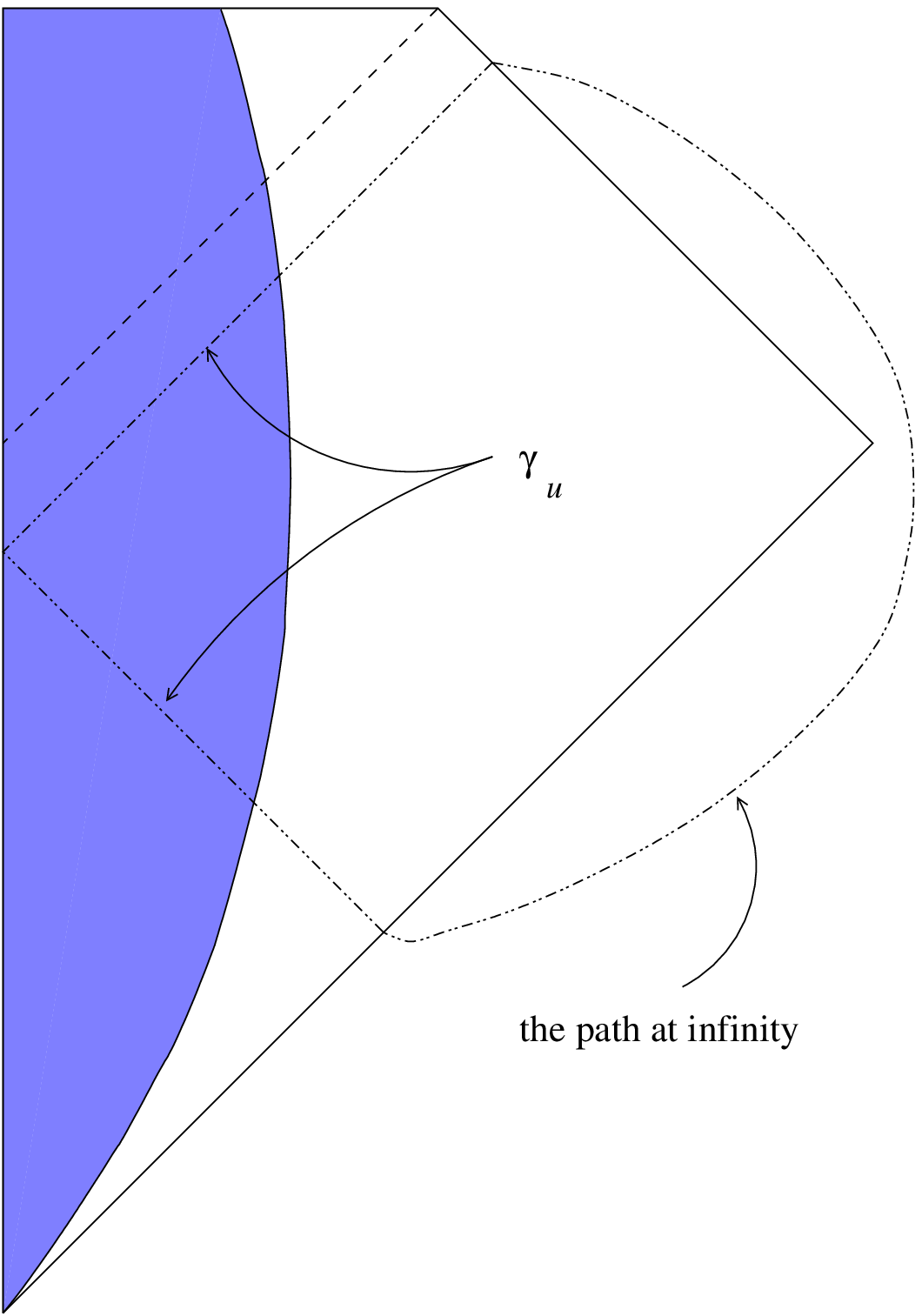}
\figure{The path $\Gamma _u$ for defining the holonomy is the geodesic
$\gamma _u$ `closed by a path at infinity'.}

\itemitem{} (a)
Fix some angular coordinates $(\theta _0,\varphi _0)$, and for each $u$ 
follow the radial null geodesic $\gamma _u(s)$ inwards from
$(u,\theta _0,\varphi _0 )$ through the space--time and out to $(v,\theta
_1,\varphi _1)$ on $\scri ^-$, where $v=v(u)$, with $v(u)$ the mapping of
surfaces of constant phase.

\itemitem{} (b) Close the path ``at infinity.''  What is really meant by this is
that we use the parallelism at infinity to identify the tangent space at
$(v,\theta _1,\varphi _1)$ with that at $(u,\theta _0,\varphi _0)$.  We abuse
terminology slightly by imagining that this parallelism arises from transport
along some mathematically fictitious path at infinity.

\noindent Transport $l^a$ and $n^a$ parallel along $\gamma _u(s)$.

Let $H^a{}_b(u)$ be the holonomy around the path $\Gamma _u$, or rather the
part of the holonomy acting on vectors orthogonal to the surfaces of
spherical symmetry, that is, those spanned by $l^a$ and $n^a$.  Since parallel
transport must preserve null vectors, the vectors $l^a$ and $n^a$ must return to
multiples of themselves, or of each other.  The latter possibility would
require a non--trivial topology and will be ignored.  Thus we must have
$$H^a{}_b =hl^an_b+h^{-1}n^al_b\eek$$
for a scalar function $h(u)$ which characterizes the holonomy.

We now derive a differential equation for $h$.  The relative increment in the
holonomy as we move from $\Gamma _u$ to $\Gamma _{u+\delta u}$ is $\delta u$
times
$$H^{-1}\dot H \ =\dot H H^{-1}
  =(\dot h /h)\left( l^an_b -n^al_b\right)\, .\eek$$\xdef\bush{\the\EEK}%
On the other hand, this quantity may be computed by integrating the curvature
tensor along $\gamma _u$.  If we parallel transport $l^a$ and $n^a$ along
$\gamma _u$, then $l^a$ will be tangent to $\gamma _u$ and $n^a\delta u$ will
be a connecting vector to $\gamma _{u+\delta u}$, so we shall have
$$H^{-1}\dot H =\int R_{pqb}{}^a l^pn^q\, \d s\, \eek$$\xdef\blair{\the\EEK}%
where $s$ is an affine parameter normalized to $l^a\nabla _as=1$ (and we
understand that the integration implicitly relies on parallel transport to
identify tensor fields along $\gamma _u$).  Comparing (\bush ) and (\blair ),
we have
$$\dot h/h=\int _{\gamma _u} R_{pqb}{}^al^pn^qn_al^b\, \d s\, .\eek$$
We next recast this as a differential equation for $v(u)$.

By definition, the result of transporting the vector $n^a=\partial _u$ at
$\scri ^+$ backwards along $\gamma _u$ to $\scri ^-$ is a vector $\dot v
(u)\partial _v$ at $\scri ^-$.  If this transport is then closed ``at
infinity,'' we come back to $\dot v (u)\partial _u$, and thus
$$\dot v (u)n^a=H^{-1}n^a=hn^a\, ,\eek$$
or 
$$\dot v (u)=h\, .\eek$$
Note that this means $\dot v (u)>0$ for all $u$.
Thus we have
$$\left(\log \dot v (u)\right)\dot{} =\int _{\gamma _u} R_D\, \d s\, ,
  \eek$$\xdef\bajpai{\the\EEK}%
where $R_D=R_{abcd}l^an^bl^cn^d$ is the sectional curvature.  This is the first
differential equation we have been seeking.  

\subsection{Interpretation and Geometry:  Evolution of $\d s$}

The equation (\bajpai ) evidently governs the evolution of the red--shift, and
so at first one would think that this is all that is necessary for
understanding the approach to the black--hole state.  However, as mentioned
above, there is a subtlety.  To understand this, note the following:

\itemitem{} (a) The sectional curvature $R_D$ is independent of the choices of
scale of $l^a$ and $n^a$ as long as the normalization $l_an^a=1$ is
preserved.   Thus $R_D$ is well--defined by the local geometry of the
space--time, and independent of issues of normalization at infinity.

\itemitem{} (b) On the other hand, the measure $\d s$ is normalized by the
requirement $l^a\nabla _as\Bigr| _{\scri ^+}=1$.  In other words, as one--forms
on the geodesic $\gamma _u$, we have $\d s=n_a$ at $\scri ^+$.  Notice this
means that $\d s$ is \it not \rm normalized relative to the Bondi frame at
$\scri ^-$, rather it has been red--shifted.

\itemitem{} (c) The measure $\d s$ is dynamic; it evolves as $u$ increases.  

The key feature of the geometry turns out to be (c).  We shall see that as the
geodesic $\gamma _u$ approaches the event horizon, the measure $\d s$ tends
exponentially to zero except on the portion of $\gamma _u$ connecting $\scri
^+$ to the horizon.  It is this, coupled with equation (\bajpai ), which gives
rise to the universality of the collapse, for it suppresses all contributions
of the geometry from times prior to the formation of the hole.

We may work out the dynamics of $\d s$ easily enough.  Let us put
$$\eqalign{n^b\nabla _bl^a&=c_l l^a\cr n^b\nabla _bn^b&=c_n
n^a\, .\cr}\eek$$
(Here $c_l$, $c_n$ would be written $2\Re\gamma$, $2\Re\gamma '$ in the
Newman--Penrose formalism.  However, we shall reserve $\gamma$ for the 
geodesic.)
Then short computations give
$$l^a\nabla_ac_l =R_D=-l^a\nabla _ac_n\, ,\eek$$
and so
$$c_l(\gamma _u(s)) =-c_n(\gamma _u(s))=
  -\int _{\gamma _u(s)}^{\scri ^+}R_D(\gamma
_u(s'))\, \d s'\, .\eek$$\xdef\braggart{\the\EEK}%
The evolution of $\d s$ is given by
$${\cal L}_n\d s=n^b\nabla _b\d s=c_n\d s\, .\eek$$\xdef\bilge{\the\EEK}%

Equation (\bilge ) shows that as long as $c_n$ is negative and bounded away
from zero in a neighborhood of the event horizon, the affine measure $\d s$
along the null geodesics will tend exponentially quickly to zero as the event
horizon is approached.  Likewise, the measure $\d s$ on the portions of $\gamma
_u$ tending to the past of the event horizon will vanish exponentially quickly,
because they are related to that on the event horizon by finite
parallel--transport factors.\fnote{There is implicit here an assumption that
the geometry becomes asympotically flat at a reasonable rate in a neighborhood
of $\scri ^-$, so that the exponential decrease of $\d s$ is not compensated by
a growth of curvature in a neighborhood of $\scri ^-$.  One would certainly
expect this assumption to hold for a  space--time corresponding to an isolated
collapsing object.}

Note that the argument above shows that, as the horizon is approached, we have
$c_n\to \int _{\gamma _u}R_D\, \d s$.

\section{Analysis and Applications}

In this section, we look at the consequences of the formulas derived
above, and in particular study what happens when a black hole forms.
We first summarize the main definitions and formulas.

The function $v(u)$ is the mapping of surfaces of constant phase, so
$\dot v (u)$ is the red--shift of a wave packet propagating through the
space--time along the radial null geodesic $\gamma _u$ terminating at
$\scri ^+$ at retarded time $u$.  The evolution of the red--shift is
governed by equation~(\bajpai )
$$\left(\log \dot v (u)\right)\dot{} =\ddot v /\dot v
  =\int _{\gamma _u} R_D\, \d s\, .
  \eqno(\bajpai )$$
Here $R_D=R_{abcd}l^an^bl^cn^d$ is the sectional curvature, the null
vectors $l^a$ (tangent to the geodesic) and $n^a$ normalized by
$l_an^a=1$. The definition of $R_D$ does not depend on further
specializations of $l^a$ and $n^a$, but other quantities in general
do, and we assume that $l^a$ and $n^a$ are transported parallel along
the geodesic and are normalized at $\scri ^+$.  The measure $\d s$ is
the restriction of the one--form $n_a$ to the geodesic, and so it is
also transported parallel along the geodesic and normalized at $\scri
^+$. The evolution of $\d s$ is given by equation~(\bilge )
$${\cal L}_n\d s=n^b\nabla _b\d s=c_n\d s\, ,\eqno(\bilge )$$
where
$$c_n(\gamma _u(s))=
  \int _{\gamma _u(s)}^{\scri ^+}R_D(\gamma
_u(s'))\, \d s'\, .\eqno(\braggart (b))$$

\subsection{Incoming and Outgoing Contributions}

Our equation~(\bajpai ) for the evolution of the red--shift involves an
integral over the whole of the geodesic $\gamma _u$.  Near an event
horizon, the contributions from different portions of this are very
unequal.  Essentially, as $u$ increases, it is the portions of the
geodesic connecting the outside of the region where the black hole
will form to $\scri ^+$ which are important, the others being
suppressed in the limit.  (This portion may be thought of as
corresponding to Stage I of the discussion in Section 2, whereas the
contribution of the earlier part of the geodesic corresponds to Stage II.)

It is therefore
convenient to re--write equation~(\bajpai ) by dividing the
integral into two parts.  We choose a point
$\gamma _u (s(u))$ on the geodesic $\gamma _u$, and divide the
geodesic into a ray $\gamma _{u+}$ outwards from $\gamma _u(s(u))$ to
$\scri ^+$ and a ray $\gamma _{u-}$ from $\scri ^-$ to $\gamma
_u(s(u))$.  (We shall discuss the choice of $\gamma _u(s(u))$
shortly.)  It will also be convenient to normalize the measure on
$\gamma _{u-}$ relative to $\scri ^-$; this measure is $\d s_-=(\dot
v)^{-1}\d s$.  We write $\d s_+=\d s$ for the measure on $\gamma _{u+}$.
Then we let
$$a_\pm (u)=\int _{\gamma_{u\pm}} R_D\, \d s_\pm\, .\eek$$
It should be emphasized that as long as there is no extreme injection
of matter from $\scri ^-$, or ejection of matter to $\scri ^+$, one
would expect the quantities $a_\pm (u)$ to remain finite and bounded
and (as long as $\gamma _u(s(u))$ does not go off to infinity)
generically non--zero.
Then equation (\bajpai ) becomes
$$\left(\log \dot v (u)\right)\dot{} 
 = a_-(u)\dot v+a_+(u)\eek$$\xdef\fred{\the\EEK}%
or equivalently
$$\ddot v = a_-(u)(\dot v)^2+ a_+(u)\dot v\,
.\eek$$\xdef\brad{\the\EEK}%

\subsection{Phase Portrait}

Equation (\brad ) can be regarded as an evolution equation for $\dot
v$.  If we temporarily ignore the $u$--dependence of $a_\pm$, then the
equation has two fixed points, one at $\dot v=0$ and the other at
$\dot v =-a_+/a_-$.  The fixed point $\dot v=0$ is attractive for
$a_+<0$ and repulsive for $a_+>0$; the other fixed point always has
the opposite character.  

In general, there are no restrictions on the signs of $a_\pm$.  The stationary
case is exceptional in that it turns out that there $a_-=a_+=0$, if $\gamma
_u(s(u))$ is chosen to be the spatial origin.\fnote{Here is an outline of the
proof. If there is a timelike Killing vector $\xi ^a$, then the world--line of
the spatial origin must be a geodesic 
with tangent $\xi ^a$, so $\xi ^b\nabla _b\xi ^a$ vanishes at
the origin.  However, a short computation by resolving $\xi ^a$ into components
along $l^a$ and $n^a$ and using Killing's equation shows that $\xi ^b\nabla
_b\xi ^a$ comes out to be $c_n=a_+$ times a non--zero spacelike vector.  Thus
$a_+=0$; by symmetry, then $a_-=0$.} This is clearly a non--representative
situation as far as the behavior governed by equation~(\brad ); this points up
the importance of studying the generic, non--stationary case.

According to the sign of $a_+$ and the position of $\dot v$ relative
to $-a_+/a_-$, then, various sorts of qualitative behavior are
possible.
(Remember that $\dot v>0$ always.)

\epsfbox{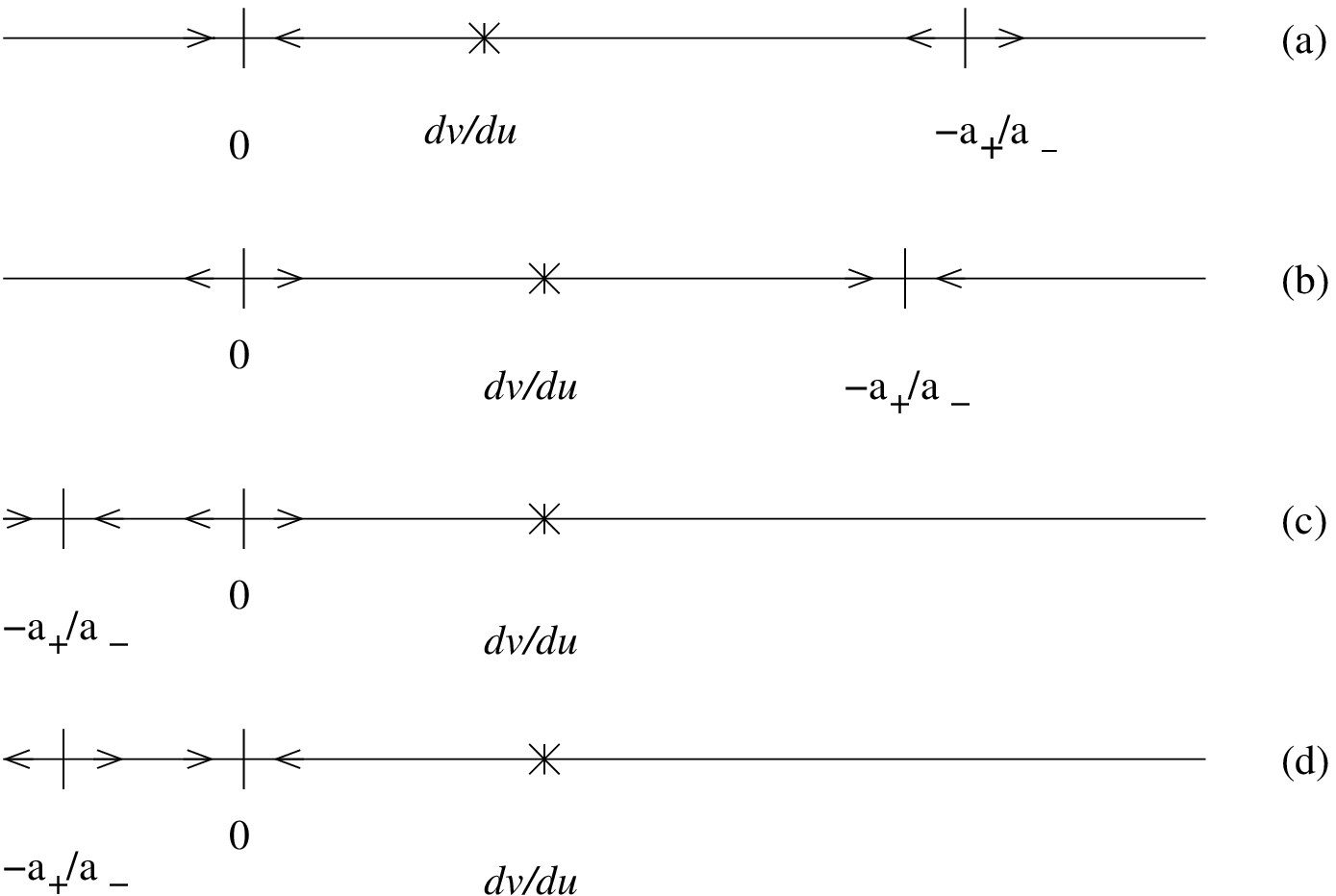}
\figure{Some possible
instantaneous phase portraits for $\dot v$.  The fixed points at zero and
$-a_+/a_-$ are indicated, with the arrows near them indicating their attractive
or repulsive character.  Cases (a) and (d) correspond to collapse, with $\dot v$
being driven to zero.  Case (b) is stable, with $\dot v$ tending to track
$-a_+/a_-$.  In case (c), one would have $\dot v$ driven towards $+\infty$.}

Approach to an event horizon means $\dot v\to 0$, and in the simplest
cases one would expect this to mean that $a_+(u)$ becomes negative and
that $\dot v a_+$ dominates $(\dot v)^2 a_-$.  Precisely, if there are
constants $a$ and $k$ such that $a_+(u)<a<0$ and (if $a_-(u)>0$)
$|\dot v a_-/a_+|<k<1$, then we shall have $\ddot v/\dot v<a(1-k)$,
and hence $\dot v\to 0$ at least as fast as $\sim\exp a(1-k)u$.
One can think of $\dot v$ as collapsing to zero in this case.  (See figure
3(a),(d).)

On the other hand, if $a_+>0$ and $a_-<0$, then $\dot v=0$ is a
repulsive fixed point and $\dot v=-a_+/a_-$ is an attractive one.
This is a stable situation.  No collapse will occur, and
$\dot v$ will tend to track the value $-a_+/a_-$.  (Figure 3(b).)

A situation with $a_+,a_->0$ would drive $\dot v$ larger; if this were
to persist it could drive $\dot v\to +\infty$ in finite retarded
time.  Indeed, by comparing equation~(\brad ) with
$$\ddot v=(\dot v)^2a_-(u)\, ,\eek$$
which is easily integrated, one sees that $\dot v\to +\infty$ at or
before the time $u$ such that
$$\int _{u_0}^ua_-(u')\d u'=1/\dot v(u_0)\, .\eek$$
It is hard to see that such a situation could be compatible with
physical requirements like global hyperbolicity and reasonable
asymptotic behavior at $\scri$.  It would mean that
by the retarded time $u$, distant
observers have already received all of the data incoming from $\scri
^-$.  (Figure 3(c).)

One can also have more complicated situations, in which the signs of
$a_\pm$ change.  Some qualitative understanding of these can be
pieced together from the observations just given, but in general one
needs to understand the functions $a_\pm (u)$ quantitatively.  For
example, we saw in the previous paragraph that $\dot v$ could be
driven to $+\infty$ in finite time.  This would mean that whatever
$a_\pm (u)$ would do beyond that would be ``too late'' to affect the
behavior of $\dot u$.

\subsection{Approach to An Event Horizon}

We have already indicated that in the simplest cases one would expect
approach to an event horizon to correspond to a situation where
$a_+(u)<a<0$ (and $|\dot v a_-/a_+|<k<1$ if $a_->0$).  Here we discuss
this situation in more detail.

(We first mention that one can verify that these inequalities
hold in a simple example.  This is constructed as follows.  One starts
with a Schwarzschild solution of mass $m_-$ and static, say perfect fluid,
interior.  Then one introduces an incoming pulse of radiation of
energy $\Delta m$ at a fixed advanced time (say $v=0$), so that one has
in effect a time--reversed Vaidya solution in the exterior of the
fluid.  If one takes $\gamma _u(s(u))$ to be the spatial origin, then
one finds
$$\eqalign{a_-(u)&=0\cr
  a_+(u)&=-{{m_+}\over{r_0^2}}
  +{{r_0-2m_+}\over{r_0-2m_-}}\cdot {{m_-}\over{r_0^2}}\, ,\cr}\eek$$
where $m_+=m_-+\Delta m$ is the total mass and $r_0$ is the value of
the radius at which $\gamma _-$ crosses $v=0$.  We have $a_+<0$.
If a black hole forms, then $r_0\to 2m_+$ and $a_+\to -1/4m_+$,
(minus) the surface gravity.)

Notice that as long as $\gamma _u(s(u))$ is chosen to continuously approach a
finite point in space--time as $u\to +\infty$, the integral $a_-(u)=\int
_{\gamma _{u-}}R_D\d s_-$ will remain bounded.\fnote{Again, this is true with
mild asymptotic assumptions at $\scri ^-$.  For example, assuming that $R_D$
has the usual peeling behavior uniformly in $v$ for a neighborhood of $\scri
^-$ near the limiting value of $v$ would be sufficient.}   Since this integral
is multiplied by $\dot v$ in equation (\fred ), it will not contribute to the
limiting behavior of $\ddot v /\dot v$.  This is true not
just for $\gamma _u(s(u))$ being, for example, the spatial origin, but even if
$\gamma _u(s(u))$ tends to any finite point on the event horizon.   This shows
that the limiting behavior of $\ddot v /\dot v$ is
independent of all details of formation of the black hole, and even of the
geometry of any finite point on the horizon.

While this behavior is easy enough to understand analytically, it is
not captured very well by any of the standard infinite regimes common
in relativity.  What is relevant here is compressed to a point in the
usual conformal diagram:  it is the ``gap'' between the event horizon
and $\scri ^+$.

\epsfbox{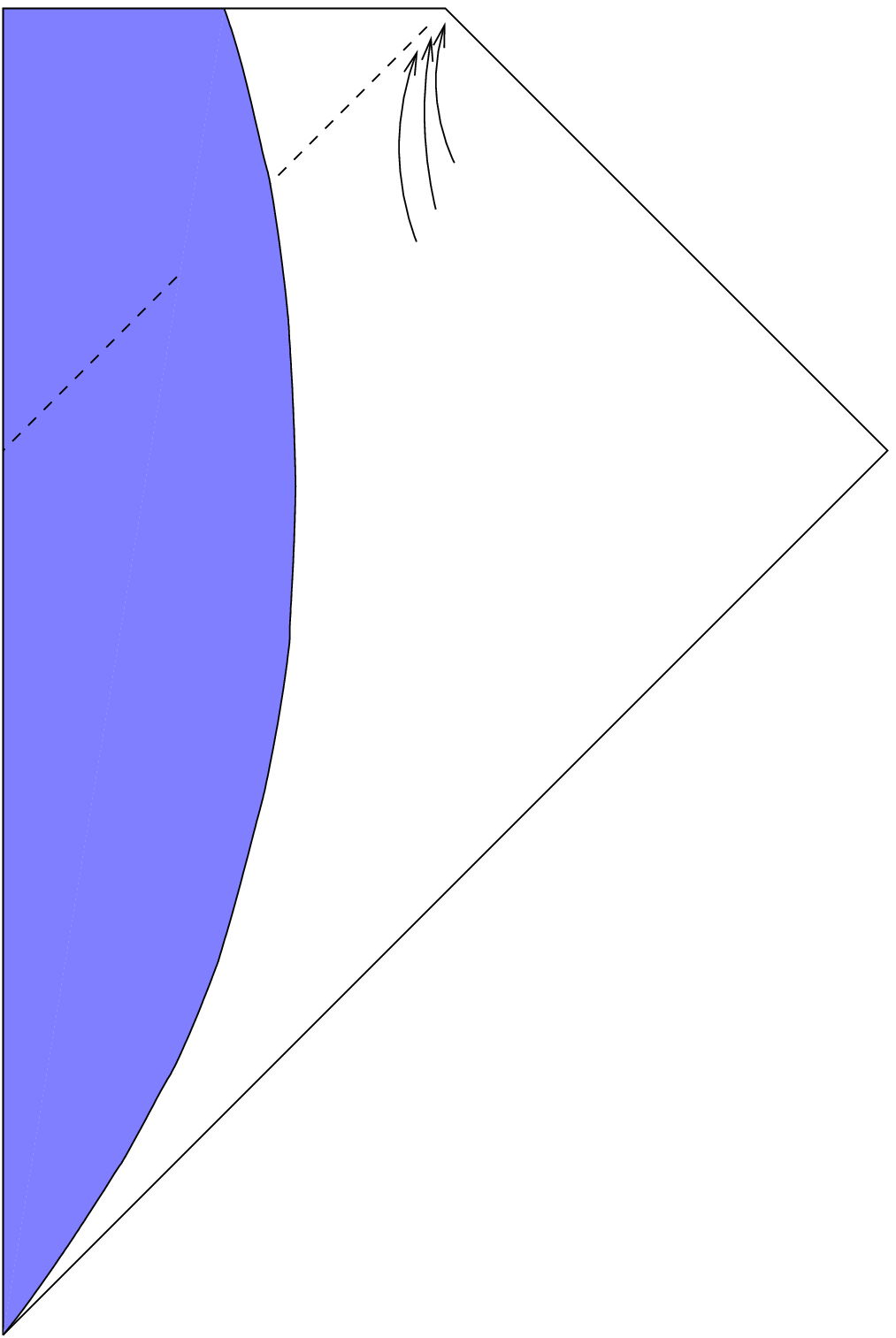}
\figure{The regime relevant to the asymptotic behavior of the fractional
acceleration of the red--shift $\ddot v /\dot v$,
indicated by arrows.  It is the gap between the event horizon and future null
infinity.  This regime is compressed to a single point in a conformal diagram,
such as this one.}

Let us recall that
the evolution of $\d s$ is given by
$${\cal L}_n\d s=n^b\nabla _b\d s=c_n\d s\, ,\eqno(\bilge )$$
where
$$c_n(\gamma _u(s))=
  \int _{\gamma _u(s)}^{\scri ^+}R_D(\gamma
_u(s'))\, \d s'\, .\eqno(\braggart (b))$$
In the situation we are considering, this approaches a constant
negative value if $\gamma _u(s)$ approaches a finite point on the
event horizon.  This means that $\d s$ vanishes exponentially quickly
as the event horizon is approached along the vector field $n^a$.\fnote{Since
$n^a$ is itself being exponentially compressed by the same factor, the measure 
$\d s$ would vanish linearly in local coordinates.}
This is really the
feature which makes the limiting behavior of 
$$\ddot v /\dot v =\int _{\gamma _u}R_D\d
s\eek$$\xdef\Frab{\the\EEK}%
independent of all details of the formation of the hole or of the
geometry of finite portions of the horizon.

Let us recall that this subsection has been concerned with the formation of an
event horizon only in the simple case $a_+(u)<a<0$, and this probably
corresponds to what happens in a ``normal'' collapse of a spherically symmetric
star.  However, it is quite possible that other, more complicated, stituations
occur, for example in critical collapse.

\subsection{Surface Gravity}

Usually, the surface gravity of a black hole is only considered to be defined
if the hole is stationary.  Indeed, the surface gravity is defined in terms of
the Killing vector $\xi ^a$ associated with the stationarity.  We shall show
here that in the case of a stationary black hole, the surface gravity is simply
(minus) $\ddot v/\dot v$, and so this quantity is a candidate for the notion of
``surface gravity'' in general.

Let us write
$$\xi ^a=\lambda l^a+\nu n^a\, ,\eek$$
for scalar functions $\lambda$ and $\nu$.  (These are not the Newman--Penrose
quantities often denoted by those symbols.)  We have $\lambda ,\nu >0$ in the
stationary region, and $\lambda ,\nu \to 1$ at $\scri ^+$ (normalization).  The
components of Killing's equation in the $l^a$, $n^a$ plane are
$$\eqalign{l^a\nabla _a\nu &=0\cr 
     c_l\lambda +n^a\nabla _a\lambda &=0\cr
     l^a\nabla _a\lambda +c_n\nu +n^a\nabla _a\nu
     &=0\cr}\eek$$\xdef\orc{\the\EEK}%
The first implies $\nu =1$ everywhere.  (The remaining system is of
course
overdetermined, giving a consistency requirement for a Killing vector to exist.)
Now, we may use these equations in any of the standard formulas for the surface
gravity.  For example, one has
$$-(1/2)\nabla _a\xi ^b\xi _b= \kappa \xi _a \eek$$
as the horizon is approached.
Contracting this with $n^a$, say, and substituting (\orc ) gives
$$\kappa =c_l=-c_n\, \quad\hbox{at the horizon.}\eek$$
We saw in the
previous subsection that this limiting behavior of $c_n$ was
$$\ddot v /\dot v =\int _{\gamma _u}R_D\d
s\, ,\eqno(\Frab )$$
and thus we identify (minus) this with the surface gravity, whether the
exterior is asymptotically stationary or not.

Two remarks are in order.  First, in the non--stationary case, one
would not usually expect $-\ddot v
/\dot v$ to tend to a constant value, and this is the reason for the
terminology ``limiting behavior'' rather than ``limiting value.''
Second, it is arguable whether this quantity ought to be called the
\it surface \rm gravity, since it depends on the global structure and
not just on the local geometry of the horizon.  (However, this is true
even when the surface gravity is defined by means of a Killing field,
for then the definition depends on the normalization of the field at
infinity.)  In any event, the identification made here of this
quantity with $-\ddot v/\dot v$ shows that it is interpretable as a
relative acceleration even in the non--stationary case; whatever
one calls it, is it the key quantity in the geometric--optics
approximation to scattering by the black hole.

\section{Comments on the Non--Spherically Symmetric Case}

Perhaps surprisingly, much of the analysis above can be extended to the
non--spherically symmetric case without great difficulty.  The
results are formulas for the gravitational acceleration of
wave--vectors by the black hole; these are no longer quantifiable by a
scalar red--shift, however.  We shall only sketch these ideas here.

Suppose that $(M,g_{ab})$ is a general space--time, which is
well--enough behaved that $\scri ^-$ and $\scri ^+$ exist as null
hypersurfaces, and $\gamma$ is a null geodesic from $\scri ^-$ to
$\scri ^+$.  (Actually, all considerations here are local to the
geodesic, so it is enough to have small portions of $\scri ^\pm$ near
the endpoints of the geodesic; one does not need much of the global
structure of $\scri$.)  

Let $l^a$ be 
a parallel propagated tangent vector to $\gamma _u$.  This will be
interpreted as the nominal wave vector of a wave packet in the geometric--optics
approximation, so we want to understand how $l^a$ is affected by its passage
from $\scri ^-$ to $\scri ^+$.  In order to do this, we must have some way of
comparing vectors on $\scri ^-$ with those on $\scri ^+$.  In general, there is
no simple way to do this, for in the presence of gravitational radiation one
does not have covariantly constant vector fields at $\scri$, and this leaves
aside the issue of relating vectors at $\scri ^-$ to those at $\scri ^+$.  

It is possible to overcome these difficulties by various technical means, but
at the present level of discussion it turns out that we do not have to address
this issue.  We suppose that a Lorentz transformation $L_u$ is chosen
identifying the tangent space at the past end--point of $\gamma _u$ with that at
the future end--point.  In general, there will be different choices of $L_u$
possible, and presumably the correct choice in any situation is dictated by the
physics.

We may then define a closed path $\Gamma _u$ as before, one portion of which is
the geodesic $\gamma _u$, and the other portion of which is a mathematical
fiction, parallel transport along the second portion being effected by
$L_u^{-1}$.  Let us write $P(u)$ for the parallel transport along
$\gamma _u$; then the holonomy around the closed path beginning and ending at
the end--point at $\scri ^-$ is $Q=L^{-1}P$.  A wave vector $l^a_{\rm in}$ at
$\scri ^-$ is taken to $l^a_{\rm out}=P^a{}_bl^b_{\rm in}$ at $\scri ^+$.  In
order to get a measure of the red shift, this must be compared with
$L^a{}_bl^b_{\rm in}$, the result of transporting the vector from $\scri ^-$ to
$\scri ^+$ by the fictional ``path at infinity.''  In the sense of this
fictional transport, and referring all vectors to $\scri ^+$, we may identify a
``red--shift operator'' $Q=PL^{-1}$.  Then we have
$$\eqalign{\dot Q &=\dot P L^{-1} -PL^{-1}\dot L L^{-1}\cr
  &=\dot P P^{-1} Q-Q\dot L L^{-1}\cr}\eek$$
In this formula, we have
$${\dot P}^a{}_c\left( P^{-1}\right) ^c{}_b =\int _{\gamma _u} R_{pqb}{}^a l^p
n^q\d s\, \eek$$
the integration of the tensor field being done by identifying the tensor field
with a tensor in the tangent space to the end--point of $\gamma _u$ on $\scri
^+$ by parallel transport along $\gamma _u$.

If $p$ is a point at which the event horizon is smooth, and near $p$ one varies
the geodesic $\gamma _u$ to approach (from the past) the generator of the event
horizon through $p$, then the normal field $n^a$ to the geodesic becomes
compressed as $p$ is approached.  This means that $n_a$, which restricts to $\d
s$ on the geodesic, tends to zero.  Thus contributions to $\dot P P^{-1}$ from
the geometry in the neighborhood of the event horizon are suppressed, and the
late--time behavior of $\dot P P^{-1}$ depends only on the portion of the
space--time connecting the event horizon to $\scri ^+$, in the same sense as for
radial symmetry.

\section{The Hawking Process}

The foregoing results have implications for the usual view of the Hawking
process.  These will be outlined here, although there are a number of
technicalities and caveats which we shall only mention, their full treatment
requiring more extensive discussion than can be given in this space.
We consider only the spherically symmetric case here.

The first point to be made is that to leading order, Hawking's model of the
emission of radiation by an object collapsing to form a black hole is isomorphic
to a moving--mirror model in two--dimensional Minkowski space.  This comes about
as follows.

The physical question at hand is:  given a massless quantum field quiescent in
the far past and propagating through the space--time of a gravitationally
collapsing body, what does the field look like at late retarded times far away
from the body?  This is a scattering problem, the object being to work out the
field operators near $\scri ^+$ in terms of those near $\scri ^-$.  

We shall restrict our attention
to the s--wave sector, which can be shown to be by far the main contribution. 
This leaves us with s--waves propagating through the collapsing object.  For
these, for those field modes of moderate wavelength near $\scri ^+$, the
geometric--optics approximation is a good one, since these arise from very
blue--shifted modes near the surface of the collapsing object and on the rest
of the space--time.  
If, as usual, we rescale the fields $\phi$ to have finite
(operator--valued) limits $\phi _0$ near $\scri$ by putting $\phi =\phi _0/r$,
then we have
$$\phi ^{\scri ^+}_0(u)=-\phi ^{\scri ^-}_0(v(u))\, ,\eek$$\xdef\greg{\the\EEK}%
where $v(u)$ is the mapping of surfaces of constant phase, and the minus sign
comes from reflection through the spatial origin.

The relation (\greg ) is formally identical to that 
for scattering of a massless field on one
side of a mirror with trajectory $v=v(u)$ (in standard null coordinates
$u=t-x$, $v=t+x$) in two--dimensional \it Minkowski \rm space.  To check that
the quantum field theories are in fact the same, one has to verify that
canonical commutation relations and field representations are the same; this is
routine, using the fact that the original space--time was asymptotically flat
near $\scri ^-$.

We can immediately write down the expectation of the stress--energy
near $\scri ^+$
from knowledge of the corresponding result in the moving--mirror theory
(cf. Birrell and Davies 1982):
$$\langle {\widehat T}_{ab}^{\rm ren}\rangle =(12\pi r^2)^{-1} \left[ {3\over
4}\left({{\ddot v}\over{\dot v}}\right) ^2-{1\over 2}{{v^{(3)}}\over{\dot
v}}\right] l_al_b\eek$$\xdef\gherkin{\the\EEK}%
for the s--wave contribution to the stress--energy near $\scri ^+$.  
Of course, this equation is only valid insofar as:  (a) we may neglect all
contributions other than the s--wave one; and (b) the geometric--optics
approximation (here expressed by equation (\greg )) is legitimate. 
The usual view is that these do give a good treatment of the leading physics. 
See Page (1976a, 1976b, 1977) for the validity of the s--wave approximation. 
See Hawking (1975, pp. 210--211) for an argument that the inclusion of 
backscattering only
alters the result (\gherkin ) by a geometric factor independent of the details 
of the formation of the hole; for another approach to this issue, see
Fredenhagen and Haag (1990).

If there is an event horizon with surface gravity
$\kappa =-\ddot v /\dot v$, then at late retarded times (and for $r\gg m$, the
mass of the object),
we will 
have
$$\langle {\widehat T}_{ab}^{\rm ren}\rangle =(12\pi r^2)^{-1} \left[ {1\over
4}\kappa ^2+{1\over 2} \dot\kappa\right] l_al_b\,
.\eek$$\xdef\vogt{\the\EEK}%
In particular, if $\dot\kappa\to 0$, or even if the time over which $\kappa$
changes significantly is much smaller that $\kappa ^{-1}$, we have
$$\langle {\widehat T}_{ab}^{\rm ren}\rangle \simeq
  (48\pi r^2)^{-1} \kappa ^2l_al_b\, ,\eek$$
which is the standard result for asymptotically stationary black
holes. 
But the
formula (\vogt ) is valid even for non--stationary black holes,
insofar as the general framework of Hawking's analysis is valid.

Several comments about this are in order:  

(a) Equation (\vogt ) can 
be viewed as giving us corrections to the
adiabatic approximation for the variation of the emission rate with the surface
gravity.  One would expect that approximation to break down when the time scale
over which $\kappa$ changes to become comparable to $\kappa ^{-1}$, and that is
just what appears in (\vogt ).  

(b) In particular, there would be no significant
modification of Hawking's prediction in the case of an isolated Schwarzschild
black hole, since in this case $|\dot\kappa |\ll\kappa ^2$ until the hole is of
Planck dimensions, when any semiclassical treatment is dubious.

(c) It is possible for $\dot\kappa$ to
be negative; indeed, one would expect $\dot\kappa <0$ for an accreting
black hole.  

(d) It is possible that in some cases $(1/4)\kappa
^2+(1/2)\dot\kappa <0$.  In this case $\langle {\widehat T}_{ab}^{\rm
ren}\rangle$ would correspond to a flux of null negative--energy
particles.  Such behavior is familiar, if not wholly understood, in
the moving--mirror case.  

(e) Whatever the sign of $\dot\kappa$,
equation (\vogt ) shows that Hawking's prediction will be
substantially modified if the surface gravity changes very rapidly.

(f) We can understand the validity of the Unruh (1976) vacuum for computing the
Hawking effect, in terms of the universality discussed above.  The point is
that the Hawking effect arises from the singular behavior of $\dot v$, and
changing the early part of the space--time in any way which only prepends a
non--singular transmission of null rays will not affect this.  Unruh's choice
of vacuum is simply such a choice which is mathematically natural.

(g) Finally, one of the most important lessons of the moving--mirror models is
that it is not possible to understand their energy budget without taking into
account the energy of the mirror and its driving engine as quantum operators,
nor without taking into account the entanglement of the field state with that
of the mirror and driving engine (Parentani 1996, Helfer 2001).  The
entanglement occurs because the field energies depend on the mirror and its
driving engine, which are themselves quantum systems:  the field energies are
functionals of the acceleration of the mirror, which is a certain quantum
operator.  This quantum character cannot be ignored for understanding the
system's energetics, because observation of the field energy requires driving
the mirror/engine system into a superposition of energy eigenstates.  If there
is a parallel with black holes, then the treatment of the collapsing object and
its space--time geometry as classical is questionable.

\references

\refbk{Birrell, N D and Davies, P C W 1982}{Quantum fields in curved
space}{(Cambridge:  University Press)}


\refjl{Fredenhagen, K and Haag, R 1990}{\CMP}{127}{273}

\refjl{Hawking, S W 1974}{Nature}{248}{30}

\refjl{Hawking, S W 1975}{\CMP}{43}{199--220}

\refjl{Helfer, A D 2001}{Phys Rev}{D63}{025016}

\refjl{Page, D N 1976a}{Phys Rev}{D13}{198}

\refjl{Page, D N 1976b}{Phys Rev}{D14}{3260}

\refjl{Page, D N 1977}{Phys Rev}{D16}{2402}

\refjl{Parentani, R 1996}{Nucl Phys}{B465}{4526}

\refbk{Penrose R and Rindler W 1984--6}{Spinors and
Space--Time}{(Cambridge:  University Press)}

\refjl{Unruh, W G 1976}{Phys Rev}{D14}{870}

%


\bye